\documentclass[a4paper,twocolumn]{esapub} % European paper
\usepackage{times}
\usepackage{natbib}
\usepackage{graphicx}

\title{The Crab nebula: Linking MeV synchrotron and 50 TeV inverse Compton photons}
\author{D. Horns}
\author{F.A. Aharonian}
\affil{Max-Planck-Institut f\"ur Kernphysik, D-69117 Heidelberg,
Germany}

\begin{document}
\maketitle
\begin{abstract}
   Pulsar wind driven synchrotron nebulae are offering a unique view on the
connection of the pulsar wind and the surrounding medium. The Crab nebula is
particularly well suited for detailed studies of the different emission
regions.  As inferred from the observed synchrotron emission  extending beyond
MeV energies, the Crab is a unique and extreme accelerator. In the framework of the
synchrotron/inverse Compton emission model, the same electrons with energies
exceeding $10^{15}$ eV that are responsible for the MeV synchrotron emission
produce via inverse Compton scattering 10-50~TeV radiation which has recently
been observed with the HEGRA system of ground based gamma-ray telescopes.  Here we discuss
the close relation of the two energy bands covered by INTEGRAL and ground based
gamma-ray telescopes. Despite the lack of sufficient spatial resolution in both
bands to resolve the emission region, correlation of the flux measurements in
the two energy bands would allow to constrain the structure of the emission
region. The emission region is expected to be a very compact region (limited by
the life-time of the electrons) near the termination shock of the pulsar wind.
We extend previous model calculations for the nebula's emission to include an additional
compact non-thermal emission region recently detected at mm wavelengths. The overall good agreement
of this model with data constrains additional emission processes (ions in the wind, inverse
Compton from the unshocked wind) to be of little relevance.
\end{abstract}

\section{Introduction}
% General features
 Observations of the Crab pulsar and nebula have been carried out in every
observationally (ground and space based) accessible wavelength band. The
resulting broad band spectral energy distribution (see Fig.~1) is exceptionally
complete and unique amongst all astronomical objects observed and studied. The
historical lightcurve of the super nova explosion of 1054 A.D. is not
conclusive with respect to the type of progenitor star  that exploded and 
leaves many questions unanswered to the stellar evolution of the progenitor star. At the
present date, the observed remnant is of a plerionic type with a bright
continuum nebula emission and filamentary structures emitting mainly in lines
in the near infrared and optical.  

% Power available
 The remaining compact central object is a pulsar with a period of 33~ms and a
spin-down luminosity of $\propto \dot P^3/P=5\cdot 10^{38}$~erg/s.  The emitted
power of the continuum peaks in the hard UV/ soft X-ray at $\approx
10^{37}$~erg/s (assuming a distance of 2~kpc).  Given the kinetic energy of the
spinning pulsar as the only available source of energy in the system, the
spin-down luminosity is efficiently converted into radiation. There is no
evidence for accretion onto the compact object as an alternative mechanism to
feed energy into the system. 

%SED
 The observed SED of the nebula is commonly interpreted as synchrotron emission
from relativistic electrons.  The general features of the SED are characterised
by 3 breaks connecting 4 power-law type spectra. The breaks occur in the near
infrared, near UV, and hard  X-ray. The break frequencies are inferred indirectly 
because they occur in spectral bands which are difficult to observe.   The high 
energy part of the synchrotron spectrum cuts of at a few MeV (see Fig.~1). This
energy is very close to the theoretical limit for  synchrotron emission derived from
balancing energy gain in diffusive shock acceleration 
and energy losses  to
be $h\nu_{max}\approx m_e\,c^2/\alpha=68$~MeV (independent of the magnetic field). 
In this sense, the Crab accelerates particle close to the extreme limit of diffusive
shock acceleration. However, the interpretation of the multi MeV emission 
as synchrotron emission requires the presence of electrons up to PeV energies. 
These 0.1-1 PeV electrons inevitably radiate via inverse Compton scattering at photon 
energies above 50~TeV with detectable fluxes
\citep{ICRC2003,1998ApJ...492L..33T,2004A&A}. Another independent indication for a common origin of
MeV and TeV photons is correlated variability of the
nebula emission in these two energy bands which will  be discussed in Sect.~4.

%Morphology
 The spatially resolved morphology of the nebula is of complex structure and
shows a general decrease of the size of the emission region with increasing
energy.  However, there are indications that at mm-wavelengths, a compact,
possibly non-thermal emission region is present that does not follow the
general behaviour \citep{2002A&A...386.1044B}. The central region close to the torus-like structure is also
known to be variable in time at  Radio \citep{2001ApJ...560..254B}, optical \citep{1995ApJ...448..240H}, 
and  X-ray frequencies \citep{2002ApJ...577L..49H}. The morphology
of these moving structures has lead to the commonly used term ``wisps'' \citep{1969ApJ...156..401S}. Whereas in 
the radio the measured spectrum is independent of the location in the nebula, at higher frequencies,
a general softening of the spectrum from the central region to the outer edge of the nebula is observed.
Recent high resolution spectral imaging at X-ray energies have revealed a \textit{hardening} of the spectrum
in the torus region which could be indicative of another location of acceleration \citep{2004astro.ph..3287M}. This is an
exciting prospect for future gamma-ray observations with better spatial resolution as will be discussed in Section~6.

%VHE 
 The very high energy (VHE) emission observed at GeV to TeV energies is
attributed to inverse Compton scatterings taking place between the electrons
and various soft photon fields present.  Based upon accurate measurements of
this inverse Compton component, a reliable estimate of the average magnetic
field can be derived resolving the degeneracy of magnetic field energy density
and particle number density for the synchrotron emission. Currently, the volume
averaged magnetic field is estimated to be 160~$\mu$G \citep{2000ApJ...539..317A}. 

%New measurements - new model calculation
The  VHE emission above 100~GeV is only observable with ground based gamma-ray instruments.
This type of instrument uses the atmospheric Cherenkov effect  to record images of
extended air showers with large optical telescopes.  The performance and
sensitivity of ground based gamma-ray instruments have improved considerably over the past decade
and allow to investigate in detail the high energy end of the spectrum of the Crab nebula and pulsar.
Recently, a new and detailed measurement of the VHE emission of the Crab
nebula has been published by the HEGRA collaboration \citep{ICRC2003}.  The wide
energy coverage from 500 GeV up to 50 TeV and beyond is of crucial importance
to understand the origin of the emission and to observe directly the acceleration of
particles to PeV energies.  

 In this contribution, we present new model calculations for the VHE emission
based upon a phenomenological approach incorporating an additional seed photon
distribution discovered at mm wavelengths \citep{2002A&A...386.1044B}.  
The model calculation will be briefly outlined
in the next section. Comparison with the VHE measurements and interpretation
will be given in Section 3. Finally, a discussion of expected variability at
energies beyond 10~TeV and correlation with MeV synchrotron emission observable
with INTEGRAL will be given followed by a discussion and an outlook.

\begin{figure*} \centering \includegraphics[width=0.8\linewidth]{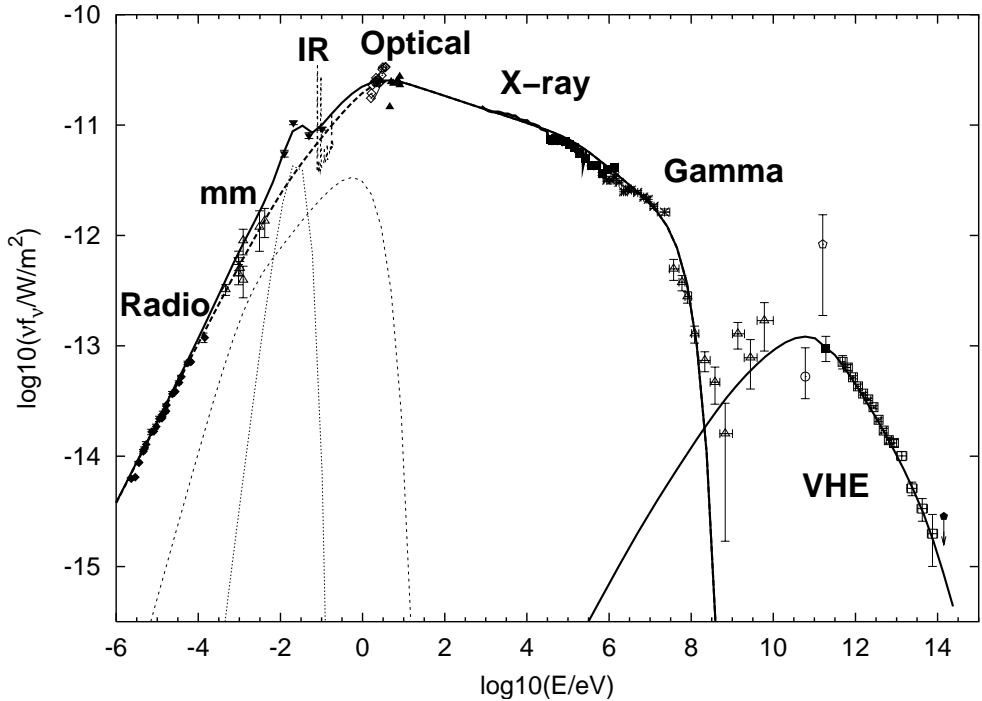}
\caption{Given the impressive coverage of different energy bands, the Crab
nebula is one of the best studied objects in multi-wavelengths.  The data used
for this compilation and their origin are described in the text.  The solid
line is the model used in this work which is the sum of the
synchrotron component (thick dashed), a thermal component (short dashed), and
the non-thermal synchrotron component at mm wavelengths (long dashed).
The compilation is not complete but chosen such that
the most recent available measurements are included.  Please refer to the text
for references and explanations on the selection of data.} \end{figure*}

\section{Model considerations}
% General model description
  With the seminal papers of \citet{1974MNRAS.167....1R} and
\citet{1984ApJ...283..694K}, a detailed picture of the evolution of accelerated
particles and the magnetic field in the nebula based upon the
magneto-hydrodynamic approximation (MHD) has evolved. In this picture, an
efficient acceleration process converts spin-down power of the pulsar into a relativistic wind injected near
the magnetosphere of the pulsar. Possible mechanisms could invoke 
electrostatic gaps which produce a dense relativistic plasma with Lorentz factors
of $\gamma\approx 10^2 -10^3$ which are then released beyond the 
light cylinder. In this picture, the energy carried by the radial outflow close
to the magnetosphere is almost entirely dominated by its Poynting flux 
(the ratio  of Poynting flux over kinetic energy flux $\sigma\approx 1000$). 

The higly relativistic but cool flow
terminates at roughly 10 arc-seconds angular distance to the pulsar when the
pressure  of the flow is balanced by the pressure of the nebula. 
Given the observation of the synchrotron emission of the nebula and 
the boundary conditions for the plasma parameters at the edge of the nebula,  
it appears that almost the entire energy of the flow 
far away from the magnetosphere is dominated by
its kinetic energy. This dramatic and so far not well understood change of the $\sigma$ parameter
is commonly referred to as the ``$\sigma$ problem''. 

The standing reverse shock quickly isotropizes the particle distribution in the
downstream frame and gives rise to a power-law type energy distribution of the
particles which then adiabatically expands and dissipates energy via 
synchrotron emission in the downstream region.
Numerous physical and observational details of this picture are still lacking a clear understanding but
this model nevertheless successfully explains the spatial and energy
distribution of electrons in the nebula responsible for the optical to hard
X-ray emission. Observations carried out with the Chandra and XMM-Newton X-ray imaging telescopes
have lead to a more detailed phenomenological picture of the distribution of particles in the nebula, which
has stipulated attempts to model the torus-type geometry with an asymmetric outflow 
\citep{2002AstL...28..373B,2003MNRAS.346..841S}. It appears that a purely toroidal field
topology is not consistent with the observations which are indicating the presence of turbulences
in the torus region.  

  In general, the calculation of the predicted VHE emission has been approached in
two different ways in the past: As a phenomenological approach, the observed
synchrotron emission has been used to extract the electron energy distribution
- with some ambiguity with respect to the magnetic field present
  \citep{1998ApJ...503..744H,1996ApJ...457..253D}.  In a more sophisticated
approach, the MHD flow model of \citet{1984ApJ...283..694K} has been used
to calculate the synchrotron and inverse Compton emissivities
\citep{1996MNRAS.278..525A} consistently.  Generally, all models give comparable results for
the predicted VHE emission.   

Besides differences in the approach which do not change the result of the calculation much, 
the calculations strongly depend on the seed photon
field(s) used to calculate the inverse Compton scattering.
Commonly, three different seed photon fields 
have been taken into account to calculate the inverse Compton emission. Here we propose a fourth one. 
\begin{itemize} 
\item
Synchrotron emission: This radiation field dominates in density for all
energies and is the most important seed photon field present. 

 \item Far infrared excess: Observations at far-infrared have shown the presence of thermal emission
which exceeds the extrapolation of the continuum emission from the radio band. This component is best
described by a single temperature of 46~K \citep{1992Natur.358..654S}. Unfortunately, the spatial structure
of the dust emission remains unresolved which introduces uncertainties for the model calculations.
 We have
assumed the dust to be distributed like the filaments with a scalelength of 1.3 arcmin. 
Sophisticated analyses of data taken with the ISO satellite indicates that the dust emission can be
resolved (Tuffs, private communication). The resulting size seems to be consistent with the value assumed here.

\item Cosmic Microwave Background (CMB):  Given the low energy of the CMB photons, scattering continues to take place
in the Thomson regime even for electron energies exceeding 100~TeV.
\end{itemize}

%It should be pointed out, that the quality of the spectral
%information available at VHE energies has improved considerably and more detailed comparisons between
%data and model are feasible (see Section 3).

Additionally, the influence of stellar light have been found to be negligible \citep{1996MNRAS.278..525A}. 
The optical line emission of the filaments is spatially too far separated from the inner region of the nebula where
the very energetic electrons are injected and cooled. However, in the case of acceleration taking place at different
places of the nebula, the line emission could be important.

 The recent claim of a detection of a compact possibly nonthermal emission region at mm wavelengths
in the center of the nebula \citep{2002A&A...386.1044B} has stipulated
us to calculate the contribution of inverse Compton scattering off of mm photons.

If the origin of the mm photons is nonthermal, it is very likely an additional
synchrotron component. The electrons responsible for this component could be
injected in a shock region closer to the pulsar - possible in the polar region.

Recent MHD calculations  carried out by \citet{2003MNRAS...344..L93} have shown
that shocks form at various distances to the pulsar. These
shocked regions could efficiently inject electrons with $\gamma=10^3-10^4$
which would produce the observed compact mm radiation  component. At the same
time, these electrons would radiate at GeV energies via inverse Compton
scattering. If the region is sufficiently compact ($<10$ arc-sec), the measured
large GeV flux by EGRET could be well explained by this component. 

It would be of great interest to investigate the spatial structure of this GeV component.
Given the size of the mm region, the corresponding inverse Compton component
would naturally be compact as well.  However, roughly half of the observed flux
is emitted by electrons which are spread over the entire nebula and would give
rise to an extended emission region.  With the advent of ground and space based
observatories with sufficiently good angular resolution and sensitivities at
roughly 10-30~GeV, the two components could be distinguished.

% New data, new problems, maybe rethink!
 Independent of the model for the injection and acceleration, it is convenient
 to derive the energy spectrum of the synchrotron emitting electrons from the
 observed continuum emission assuming an average magnetic field. The
 characteristic size of the synchrotron emitting region is derived from
 measurements and used to calculate the average photon density distribution
 that participate in the inverse Compton scattering.  The main assumption is
 spherical symmetry and that the surface brightness map follows a Gaussian
 distribution.  This approach was originally introduced by
 \citet{1998ApJ...503..744H}. The major differences with the original treatment
 are updates of the measurements, inclusion of an additional seed photon field,
 and  a modification of the electron distribution to fit the break in the near
 IR better.

 For the purpose of compiling  and selecting available
measurements in the literature, mostly recent measurements have been chosen.
The prime goal of the compilation is to cover all possible wavelengths from
radio to gamma-ray. The radio data is taken from \citep{1972A&A....17...172}
and references therein, mm data from
\citep{1986A&A...167..145M,2002A&A...386.1044B} and references therein, and the
infra-red data from \citep{2001A&A...373..281D,1992Natur.358..654S}. 

 Optical and near-UV data from the Crab nebula requires some extra
considerations. The reddening along the line-of-sight towards the Crab nebula
is a matter of some debate. For the sake of homogeneity, data in the optical
\citep{1993A&A...270..370V} and near-UV and UV \citep{1992ApJ...395L..13H,
1981ApJ...245..581W} have been corrected applying an average extinction curve
for $R=3.1$ and $E(B-V)=0.51$ \citep{1989ApJ...345..245C}. This consistent treatment
ensures that the data of different publications can be combined more easily.

 The high energy measurements of the Crab nebula have been summarized recently
in \citet{2001A&A...378..918K} to the extent to include ROSAT HRI, BeppoSAX
LECS, MECS, and PDS, COMPTEL, and EGRET measurements covering the range from
soft X-rays up to gamma-ray emission. For the intermediate range of hard X-rays
and soft gamma-rays, data from Earth occultation technique with the BATSE
instrument have been included \citep{2003ApJ...598..334L}. 

 The available VHE emission data is a collection of non-imaging
\citep{2002APh....17..293A,2002ApJ...566..343D,2001ApJ...547..949O} and imaging
ground based air shower experiments \citep{ICRC2003}. 

 The mm-component is modeled with an angular size of 36 arc-sec as a value 
 derived from
Fig.~5 of \citep{2002A&A...386.1044B}.  
The origin of the compact component is
very likely non-thermal  as pointed out in
\citet{2002A&A...386.1044B}. An ultra-cool ($<5$~K) dust component required to match
the data would have
to be unrealistically massive ($>100~M_\odot$). 

\section{Comparison of the model to data}

 The overall broadband spectral energy distribution is described quite well by
 the model which is a good cross-check for the procedure (see Fig.~1). However,
 the good agreement is not a surprise, because the electron distribution is chosen to
 reproduce the data.

 The prediction for the inverse Compton emission at high energies is compared
 with the data in Fig.~2. The different contributions of seed photons is
 indicated in the same picture. Obviously, the synchrotron seed photons are
 dominant at all energies. However, with increasing photon energies, the
 contribution of the CMB and notably of the mm radiation is of comparable
 importance as the synchrotron photon field. This is a consequence of the
 Klein-Nishina  suppression of the inverse Compton scattering at high center of
 momentum energies which strongly suppresses the contribution of the
 higher energy synchrotron seed photons to the inverse Compton emission.

 The overall agreement of the VHE prediction with the data is excellent. Over
 the entire range of energies above 500~GeV up to 80~TeV, the observed spectrum
 is well described by the model calculations.  In fact, ignoring the contribution of the
 compact mm emission region reproduces the data still reasonably well.

 The agreement of the model based upon an electron distribution inferred from the synchrotron emission 
 is important for constraining possible other emission mechanisms present.
 This has a number of immediate consequences for the interpretation: 

\begin{enumerate}
 \item The presence of very energetic electrons with energies well beyond 100~TeV
 is required to explain the multi-TeV emission as inverse Compton radiation. The
 same electrons would comfortably produce the observed MeV radiation via synchrotron
 emission provided that the magnetic field is O(100~$\mu$G).

 \item There is no indication for any additional components as expected 
 in the presence of ions in the wind \citep{2003A&A...402..827A}.

 \item There is no indication for any additional components as expected
 from the unshocked wind  \citep{2000MNRAS.313..504B}.
\end{enumerate}

\begin{figure*}[ht!]
\centering
\includegraphics[width=0.8\linewidth]{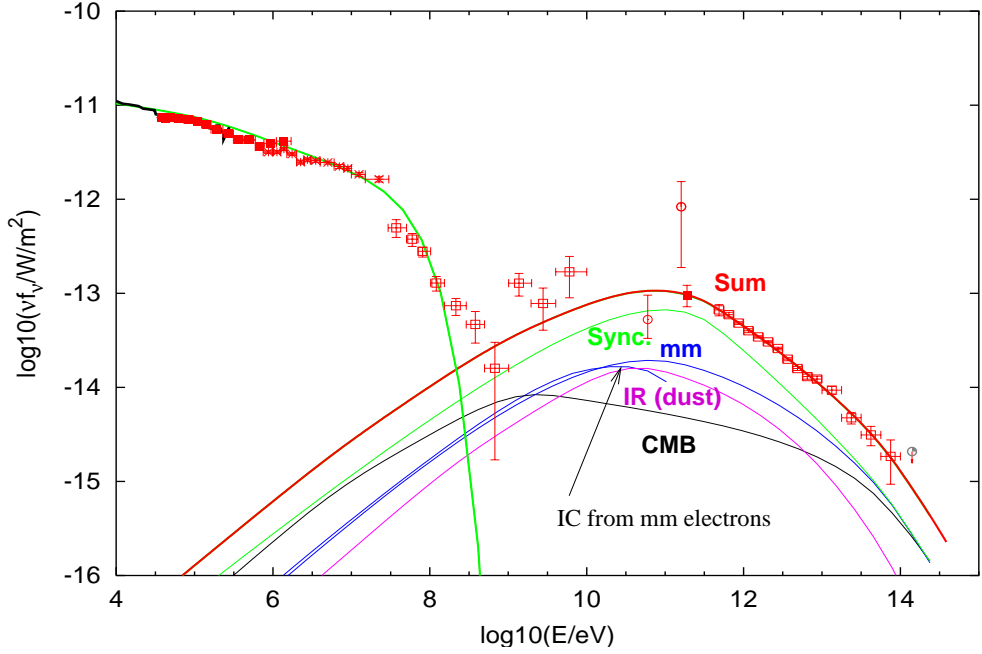}
\caption{The VHE emission decomposed into different seed photon contributions compared
with a collection of data and including the most recent VHE spectral measurement
published by the HEGRA collaboration. The dominating seed photon field is the 
synchrotron emission. At high energies, the CMB and the mm component are 
of equal importance. Assuming a synchrotron origin of the mm component, the
electrons produce GeV emission via inverse Compton scattering. The flux 
depends on the size of the mm emission region. Provided that the size is of the order
of 10 arcsec the GeV flux measured by EGRET could be well matched  by this model.
}
\end{figure*}
 
 Independent of the absolute flux predicted in the model and compared with the
 measurements, it is interesting to compare the predicted shape, namely the predicted
 gradual steepening of the spectrum with the observations. To do so, the differential
 spectra are used to calculate the slope as a function of energy. 
 \begin{eqnarray}
   \Gamma\left(\frac{\ln\nu_1+\ln\nu_2}{2}\right)        &=& \frac{\ln\Phi_1-\ln\Phi_2}{\ln\nu_1-\ln\nu_2}
   \end{eqnarray}
    Based upon Eqn.~1 with error propagation assuming that the error on $\nu$ $\sigma_\nu=0$, an estimate
    on the error of the photon index can be derived.
   \begin{eqnarray}
   \sigma_\Gamma &=& \frac{1}{\ln\nu_1-\ln\nu_2}
{\sqrt{      \left(\frac{\sigma_{\Phi_1}}{\Phi_1}\right)^2+
   							          \left(\frac{\sigma_{\Phi_2}}{\Phi_2}\right)^2}} 
								  	 \end{eqnarray}

 For the data, we have
 chosen to calculate the logarithmic slope between data points separated by 0.625 in 
 decadic logarithm.
  This is a compromise between the required leverage ($\ln\nu_1-\ln\nu_2$) which should be
 sufficiently large to keep the statistical error small and still retain the  sensitivity for
 possible changes in the slope.  The resulting differential plot of slope is shown 
 in  Fig.~3 comparing the data with the model calculated here (solid line), the same model
 excluding the mm emission (long dashed line), and the calculation in \citet{1998ApJ...503..744H}. 

 Again, the data confirms the expected trend of the model of a gradual
 steepening. It is noteworthy, that the energy range covered by presently
 existing VHE data is sampling an energy interval which exhibits only very
 small changes in the photon index. With the existing data, a pure power law
 fit is still acceptable in terms of $\chi^2$-probability. However, looking at
 Fig.~3, it is immediately clear, that the systematic trend of the data is not
 following a random pattern.  The slope of the model spectrum gets slightly
 steeper with increasing energy. Ignoring an additional contribution from a
 compact mm emitting region softens the spectrum slightly. Given the
 statistical uncertainties, both possibilities are consistent with the data. In
 any case, even with the minimal assumption of the CMB field and the synchrotron 
 photons to participate in the inverse Compton scattering, the VHE flux is
 well explained by electronic emission with a slightly lower magnetic field of $140~\mu$G.

\begin{figure}[ht!]
\includegraphics[width=\linewidth]{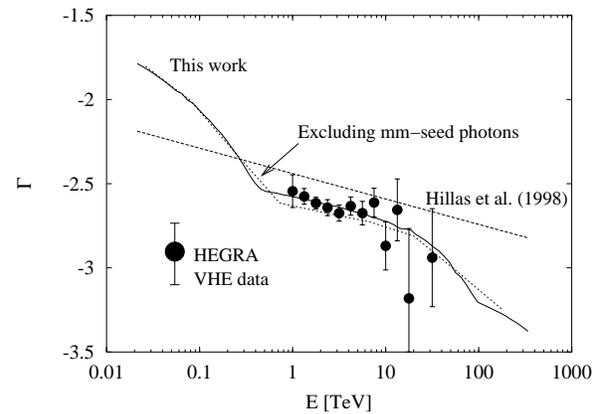}
\caption{As a function of energy, the model presented here gives a distinct prediction
for the change of the photon index. The gradual steepening of the energy spectrum in the
energy range above a few hundred GeV is confirmed by the data. Note, the dashed line
indicates the change of slope as it has been proposed in \citet{1998ApJ...503..744H}.} 
\end{figure}

\section{Correlated variability: Linking the MeV and TeV emission}
Together with the advent of INTEGRAL and its sensitivity up to MeV energies, it
is becoming feasible to study correlated time variations in the MeV and TeV
energy band. The MeV photons are emitted by electrons with typical energies
exceeding 100~TeV in the average magnetic field ($\gamma=2\cdot 10^8
(\epsilon/400\mathrm{keV})^{1/2} (B/160~\mu\mathrm{G})^{-1/2}$). 

Given the
fast decline of the cross section for inverse Compton scattering at center of
momentum energies $s^{1/2}\gg mc^2$, these electrons scatter predominantly on
mm wavelength radiation producing roughly photons at energies exceeding 10~TeV. 
The lifetime of electrons at these energies is of the order of a few months. It seems natural to
expect variations of the synchrotron flux at hundreds of keV and in fact, 
indications of variability have been claimed \citep{2003ApJ...598..334L}.

Given the short lifetime of the electrons, the emission region is expected to
be  spatially compact and  unresolved by gamma-ray instruments in the near future. 
However, by finding a correlation between the two bands, it is possible to establish
\textit{directly} that the same population of electrons is responsible for the emission via
synchrotron and inverse Compton scattering and constrain the size of the
acceleration/emission region by measuring the typical timescale of variability.

\section{Summary and Conclusions}

 The observations of MeV synchrotron and multi-TeV inverse Compton emission from the Crab nebula
 establishes the presence of energetic electrons reaching PeV energies. The mechanism responsible for
 accelerating electrons to PeV energies needs to be efficient and fast. The highest obtainable energy for
 synchrotron photons in the framework of diffusive shock acceleration in the Bohm limit is independent of
 the magnetic field strength and approximately 70~MeV. Given the observed steep cutoff in the spectral
 energy distribution of the Crab nebula at a few MeV, the accelerating mechanism present is within an order of 
 magnitude to the theoretical limit. The recent detection of gamma-ray emission at energies exceeding 50~TeV
 is consistent with the expectation for inverse Compton emission. At energies beyond 50~TeV, soft seed 
 photons at mm wavelengths are of similar importance as the synchrotron seed photons. The presence of ions 
 in the wind is not evident from the comparison of the data with the model. 
 In some models,  ions in the wind are invoked to explain the
 surprisingly high efficiency of acceleration of electrons (positrons) in the shock \citep{1992ApJ...390..454H}.  These ions
 should carry a substantial amount of the total energy flux of the wind. In principle, the unshocked wind of electrons
 would give rise to bulk Comptonization of the intense photon field emitted from the pulsar's surface \citep{2000MNRAS.313..504B}.
 Again, the good agreement of data and the model presented here, leave no ``room'' for these additional components unless
 the wind's Lorentz factor exceeds $\approx 10^7$. 
  Finally, if the mm emission is confirmed and its spatial structure better known, it is interesting to note that if this
  component is emitted by electrons, these electrons could naturally explain via inverse Compton scattering the excess of the GeV
  flux compared with the current model calculations.

\section{Outlook and perspective}
 The long history of observations of the Crab nebula with  subsequent surprises continues. With the advent of better
 spatial resolution at mm wavelengths an important conclusion for VHE emission can be drawn. More and better observations 
 at mm wavelengths are needed to study the structure of the shock waves in the wind termination zone. In fact, given the
 spatial spectral resolution at various wavelengths, dramatic improvement of the understanding of the particles' injection,
 transport, and cooling in the nebula is to be expected. Given the recent X-ray observations which show the hardest spectrum
 in the torus of the nebula far away from the inner ring which has been commonly associated with the site of acceleration, 
 very exciting times for gamma-ray observations are expected. In fact, if acceleration takes place at the torus, the 
 size of the VHE emitting region could be larger than originally calculated based upon the assumption that the inner ring is
 the exclusive site of acceleration. With the advent of the new ground and space based gamma-ray detectors, this structure
 might be resolvable. 

Finally, the observation of temporal correlation between the MeV and 50 TeV emission of the nebula will give
us important insight into the extreme accelerator which powers the Crab nebula. We expect that especially 
ground based Cherenkov telescopes operating in the southern hemisphere (CANGAROO, H.E.S.S.) will collect 
efficiently with large collection areas of more than 1~km$^2$ these very energetic photons. It is expected, that
the Crab will be often observed as a calibration target for INTEGRAL and ground based detectors. This database
of observations over many years will allow to study variability and correlations which are expected given the
life-time of the electrons responsible for the emission.

\section*{Acknowledgments}
The authors would like to thank the organizers of the meeting for their great work.

% The following bibliography was produced with
%   \bibliographystyle{aa}
%   \bibliography{esapub}
% The results are inserted directly here to simplify
% the demonstration.

\end{document}